\documentclass{aa}
\usepackage{graphicx}
\newcommand{\so}{ISO J1324$-$2016}
\begin{document}
\title{The first ISO ERO: a dusty quasar at $z = 1.5$
\thanks{Based
on observations collected at the ISO Observatory,
Canada-France-Hawaii Telescope, Australia Telescope Compact Array,
UKIRT, ROSAT and the European Southern Observatory
(prg: 58.E-0885, 61.B-0369, 265.B-5005) }}
\author{
M. Pierre\inst{1} \and C. Lidman\inst{2} \and R. Hunstead\inst{3}
\and D. Alloin\inst{2} \and M. Casali\inst{4} \and C.
Cesarsky\inst{2} \and P. Chanial\inst{1} \and P.-A. Duc\inst{1}
\and D. Fadda\inst{1} \and H. Flores\inst{1} \and S.
Madden\inst{1} \and L. Vigroux\inst{1}}
\offprints{M. Pierre, mpierre@cea.fr}
\institute{CEA Saclay,  DSM/DAPNIA, Service d'Astrophysique,
F-91191 Gif sur Yvette \and European Southern Observatory,
 Karl-Schwarzschild-Str.\ 2, D-85748 Garching bei M\"unchen
\and School of Physics, University of Sydney, NSW 2006,
Australia \and Royal Observatory, Blackford Hill, Edinburgh, EH9
3HJ, UK }
\date{accepted}
\maketitle
\begin{abstract}
We report the discovery of an extremely red object (ERO) in a
medium-deep ISOCAM extragalactic survey. The object is also a radio
source. Subsequent VLT NIR spectroscopy revealed a prominent H$\alpha$
line giving a redshift of 1.5. We present the spectrum and photometric
data points and discuss evidence that \so\ is a quasar harbouring a
significant amount of very hot dust.
\end{abstract}
\section {Introduction}
In the course of the ISOCAM Core Programme devoted to the observation
of $z\sim 0.2$ galaxy clusters (DEEPXSRC), we have discovered a faint
field source at 7.5 and 15 $\mu$m with no obvious counterpart on
medium deep optical images. This object was thus a {\em potential\/}
`extremely red object' (ERO), but {\em for the first time}, being
directly unveiled through mid-infrared (MIR) observations. The new ERO
class, usually defined by $R-K > 5$ or $R-K > 6$, is not only rapidly
growing in size but also in astrophysical relevance. Indeed, it may
shed light on the still hotly debated question of AGN/starburst
connections, the formation epoch of ellipticals as well as the
existence of dust within the crucial redshift range $1 < z < 3$. For
an up-to-date review on the ERO topic, see for instance Liu et al.\
(2000). Here, we describe the follow-up observations we have
undertaken in order to determine the redshift and to shed light on the
nature of this peculiar object. The next section presents the
spectroscopic and photometric data from X-ray to radio
wavelengths. Section 3 discusses possible interpretations in
conjunction with information provided by galaxies and other EROs at
comparable redshift.  Throughout the paper, we assume $H_{0} = 75$
km\,s$^{-1}$\,Mpc$^{-1}$ and $q_{0}$ = 0.
\section {The data set}
{\em ISO data}\\ The ISO observations of A1732 ($z = 0.193$) were
 performed in February 1996 and preliminary results are described in
 Pierre et al.\ (1996). The data have since been reprocessed,
 following the method presented by Fadda et al.\ (2000), based on
 extensive simulations (addition of faint sources to the science
 images).  This provides the flux reconstruction factors, which are
 observation dependent, as well as error estimates, given here at the
 1$\sigma$ confidence interval. Results for source \so\ are given in
 Table 1. The flux densities are below 1 mJy, which explains why \so\
 does not appear in the IRAS Faint Source Catalogue at any
 wavelength. \so\ is identified as source 2 in Fig.\ 1 of
 Pierre et al.\ (1996).
\begin{figure*}[t]
\centering
\includegraphics[height=6cm,angle=0]{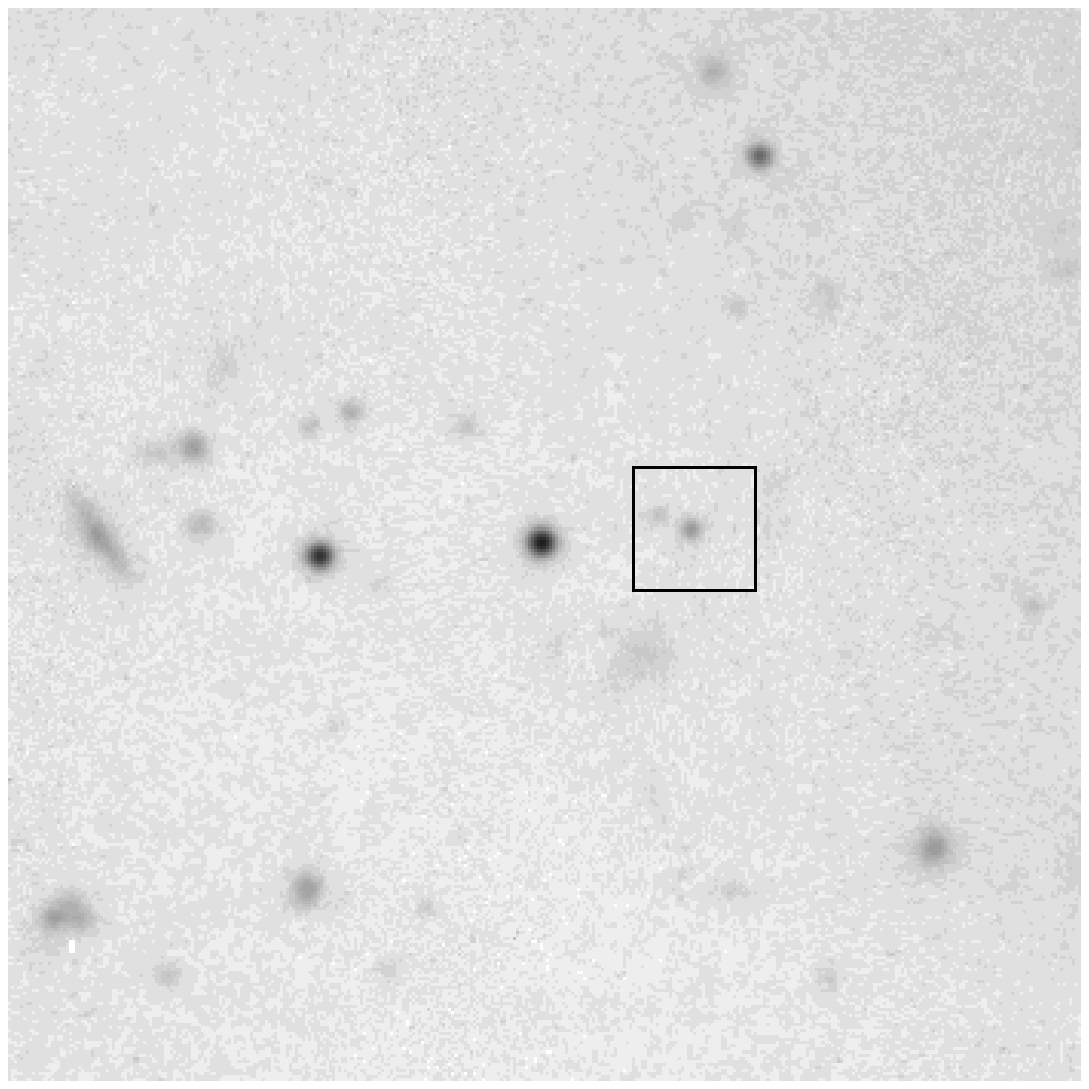}
\hspace{3cm}
\includegraphics[height=6cm,angle=0]{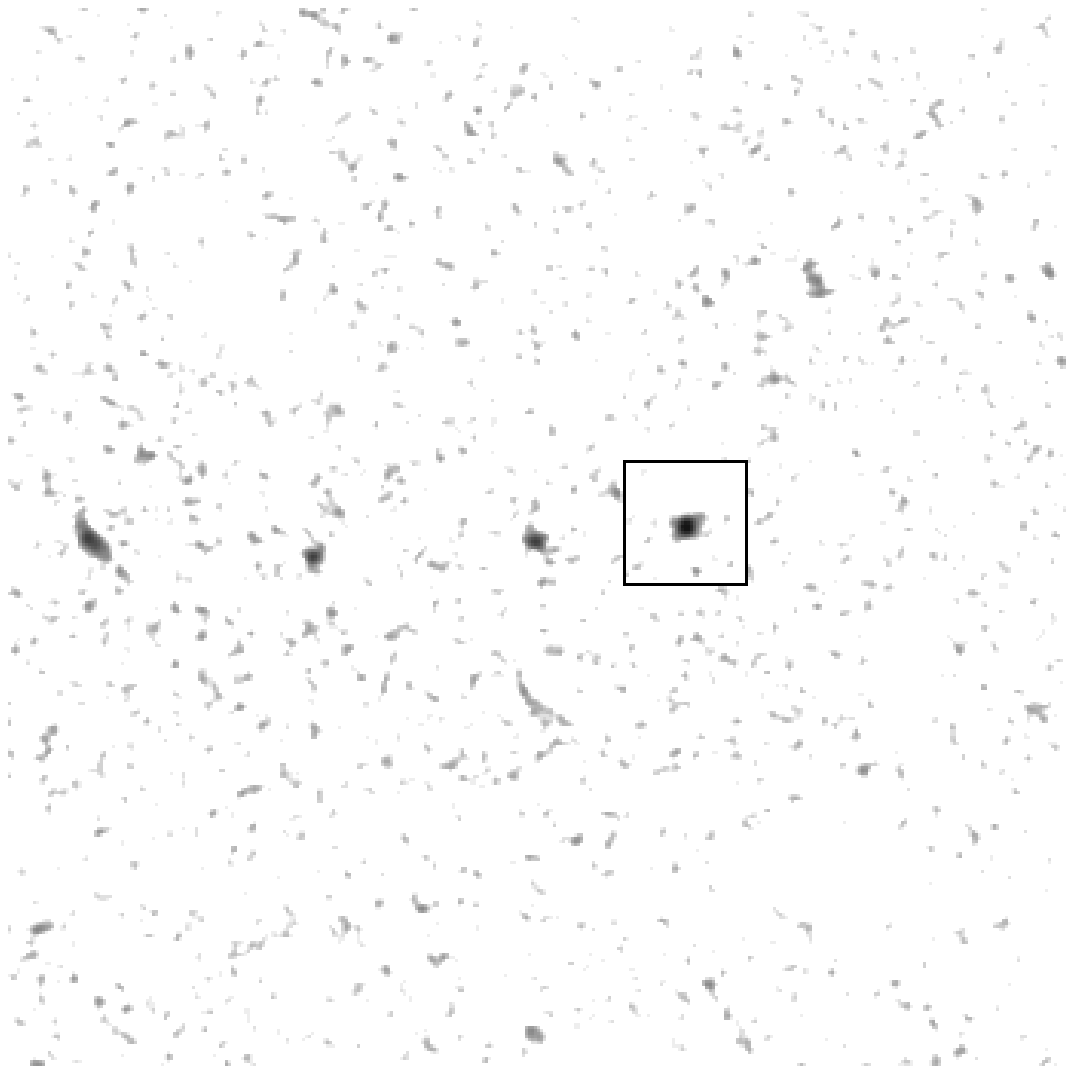}
\caption [] {Finding charts for \so. The images are $45''$ on a side;
North is up, East is left. \\ {\em Left}: ESO NTT/SUSI $I$ image
(exp. time: 1800 s). {\em Right}: ESO VLT/ISAAC $K$ acquisition
image (exp. time: 40 s)}
\end{figure*}

\noindent{\em Radio data}\\ The cluster A1732 was observed with the
 Australia Telescope Compact Array (ATCA) on 1995 April 18 at
 frequencies of 1.344 and 2.378 GHz. Total integration time was 10
 hours in the 6C array, which gives interferometer spacings from
 153~m--6~km. The synthesised half-power beamwidths at the declination
 of A1732 were $16.8'' \times 6.8''$ (PA $1.4^{\circ}$) at 1.344 GHz
 and $9.6'' \times 4.1''$ (PA $2.3^{\circ}$) at 2.378 GHz.  The
 primary flux density calibrator was PKS B1934$-$638, with B1245$-$197
 and B1622$-$297 as secondary phase calibrators.  \so\ appeared as an
 unresolved radio source, with a fitted position of 13 24 45.67 $\pm
 0.03$, $-$20 16 11.3 $\pm 0.5$ (J2000), and flux densities of $1.4
 \pm 0.12$\footnote{The value quoted in Pierre et al.\ (1996) is
 incorrect} and $0.8 \pm 0.1$ mJy at 1.344 and 2.378 GHz,
 respectively. The spectral index over this interval is $\alpha =
 -1.0$ ($S_\nu \propto \nu^{\alpha}$). The $0.5''$ radio positional
 accuracy was essential for the follow-up identification work.\\
{\em Optical/NIR broad-band imaging}\\ From 1993 CFHT images of A1732,
the following upper limits on \so\ were set: $B> 24.5$, $R >
22.7$. However, since the source was close to the CCD edge, and
affected by vignetting, these limits do not provide a useful
constraint. The source was then observed in 1997 (March 05--06) at ESO
with NTT/SUSI for two hours in the $I$ band; the seeing was
$0.6''$. In the radio error box, we discovered an object with $I =
22.4 ~ \pm$ 0.1 (see Fig.\ 1). The $I$ image of the identification is
unambiguously pointlike (FWHM $< 0.6''$). Subsequently, we obtained a
UKIRT service image (1998 April 21) of the field and measured a $K$
magnitude of 17.5. This magnitude of $K = 17.5 \pm 0.1$ was later
confirmed by the VLT/ISAAC acquisition image (see Fig.\ 1); moreover,
with a seeing of $0.4''$, the ISAAC $K$ image of \so\ remains
pointlike. The $I-K$ color of 4.9 emphasises the extreme redness of
the source spectrum --- at least in the observed frame --- and the
present lower limit, $R-K > 5.2$, reinforces the status of \so\ as an
ERO.\\
{\em X-ray observations}\\ The field of Abell 1732 was observed
with the ROSAT HRI for $\sim$ 30 ks (1996 January 15--27; Pierre
et al.\ 1996) and with ASCA for a total integration time of $\sim
$ 90 ks and $\sim $ 100 ks by the SIS and GIS instruments
respectively (1997 July 6--7; Pierre et al.\ 1999). \so\ is not
detected by the HRI. This sets a 3$\sigma$  flux upper limit of $
\sim 2\times 10^{14}$ erg~s$^{-1}$~cm$^{-2}$ in the [0.1--2.4] keV
band, assuming a standard power-law spectrum with a photon index
of 2 (or 4 $\times 10^{14}$  erg~s$^{-1}$~cm$^{-2}$ if corrected
for Galactic absorption). In the [0.4--10] keV ASCA images,
because of the large instrumental point spread function, the
cluster image encompasses the \so\ position, making its detection
impossible.\\
{\em The NIR spectroscopy } \\ A 2-hour spectroscopic observation
(1998 May 23) at intermediate resolution with EFOSC on the ESO 3.6-m
telescope over the range 4000--9000~\AA\ did not reveal any
significant absorption/emission features.
\begin{figure*}
\centerline{
\includegraphics[width=12cm,angle=0]{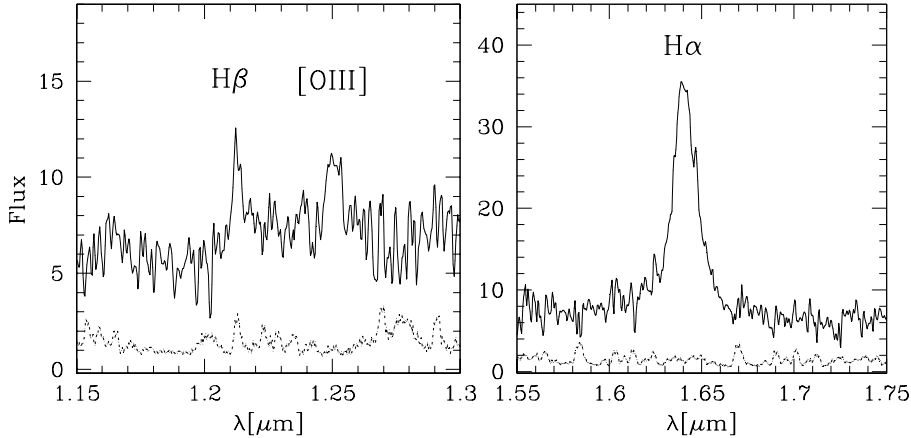}}
\caption [] {{\em Right:} Portion of the ISAAC $H$-band spectral interval
showing the H$\alpha$ line at a redshift of 1.50. The
original spectrum has been lightly filtered by a Gaussian with
a $\sigma$ of 1 pixel (3.6 \AA). Intensity is in arbitrary
units. {\em Left:} Portion of the ISAAC $J$ spectral interval
containing the H$\beta$ and [O\,III] lines. No line is detected in
the $K$-band spectrum.  In each panel, the lower curve is the noise
spectrum extracted from the same data. }
\end{figure*}
This stressed the need for deep NIR spectroscopy with a larger
telescope. \so\ was observed with the low resolution spectroscopic
mode of VLT1/ISAAC on 2000 June 7 \& 9. Three grating settings
were used to cover the 1.1--2.5 $\mu$m range. As is standard in
the IR, the target was observed at two positions along the slit
(which we shall call the A and B beams). The bright and variable
night sky lines were removed by subtracting the respective spectra
from each other. The resulting two-dimensional spectra were then
corrected for slit distortion and wavelength calibrated with the
OH lines or with arc lamps.
Residual lines from the night sky were then removed by combining
spectra from the A and B beams. This process works well enough that
one-dimensional spectra can be extracted without any need for
additional sky subtraction. In addition to \so, two hot stars, with
spectral type A0 or earlier, were observed with the same instrument
configuration. These stars were used to remove telluric features in
the spectra of \so.
The one-dimensional spectra are shown in Fig.\ 2; spectral resolution
is 21.4 and 28.4 \AA\ in the $J$ and $H$ bands respectively. The
redshift from the H$\alpha$ line is $1.500 \pm 0.002$. The H$\beta$
and [O\,III] lines show a relative blueshift of some 1500 and 800
km~s$^{-1}$ respectively. The Balmer decrement is uncertain
because the H$\beta$ line is very noisy (due to a sky line at 1.215
$\mu$m) and appears to be narrower than H$\alpha$; the estimated value of
$20^{+10}_{-4}$ is significantly higher than more common values of
5--10. The total corresponding $B$ extinction would be $A_{B} =
7^{+2}_{-1}$ in the source restframe, assuming standard extinction
laws (Mathis 1990).
\begin{table*}
\caption[]{Photometry of \so. $B,~ R,~ I~ \&~ K$ flux densities
have been corrected for galactic extinction (NED).  }
\begin{center}
\begin{tabular}{|lrrc|} \hline
Observed & \multicolumn{1}{c}{Rest} & \multicolumn{1}{c}{Flux} & Telescope/ \\
Wavelength & \multicolumn{1}{c}{Wavelength} &
\multicolumn{1}{c}{Density} & Instrument\\ \hline
 B$_{J}$ ~~ 0.44 $\mu$m & 0.176 $\mu$m& $<0.98$ $\mu$Jy  & CFHT (1993) \\
 R$_{J}$ ~~ 0.70 $\mu$m& 0.28 $\mu$m& $<2.8$ $\mu$Jy  & CFHT (1993) \\
 I$_{J}$ ~~ 0.90 $\mu$m& 0.36 $\mu$m& 3.0 $\pm ~ 0.3$ $\mu$Jy  & NTT/SUSI (1997) \\
 K$_{J}$ ~~ 2.2 $\mu$m& 0.88 $\mu$m& 67 $\pm ~6$ $\mu$Jy  & UKIRT (1998) \& VLT/ISAAC (2000)\\
 LW2 6.75 [5--8.5] $\mu$m &  2.7 $\mu$m&0.89$_{-0.33}^{+0.47}$ mJy & ISOCAM (1996) \\
 LW3 15 [12--18] $\mu$m & 6.0 $\mu$m &0.76$_{-0.40}^{+0.87}$ mJy& ISOCAM (1996)\\
 13 cm & 5.0 cm& 0.8 $\pm$ 0.10 mJy & ATCA (1995) \\
 22 cm & 8.9 cm& 1.4 $\pm$ 0.12 mJy& ATCA (1995)\\
 35 cm & 14.2 cm& $<$2.1 mJy & MOST (1993) \\ \hline
\end{tabular}
\end{center}
\end{table*}
\section{Discussion}
We note first that, although \so\ is in the field of A1732, it is
certainly not a lensed object (at least in the strong regime), since
it is located 0.8 Mpc ($4.8'$) in projected distance from the cluster
centre. The VLT/ISAAC spectrum unambiguously demonstrates that \so\ is
powered by an active nucleus, a fact already suggested by its
pointlike appearance in the $I$ and $K$ bands, together with its radio
activity. The H$\alpha$ FWHM ($\sim 3000$ km~s$^{-1}$) and rest frame
equivalent width (300 \AA) are very similar to those seen in other
high redshift quasars (Espey et al.\ 1989).
\begin{figure}
\includegraphics[width=14cm,angle=0]{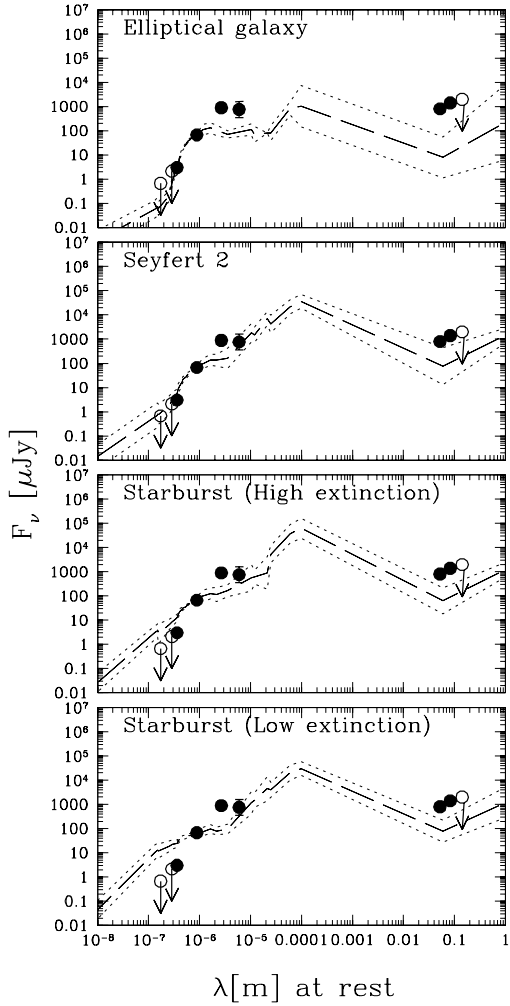}
\caption [] {Rest frame spectral energy distribution (SED) of \so\
(filled circles) compared with typical local galaxy SEDs from Schmitt
et al.\ (1997). The template spectra have been normalised to the $K$
magnitude point. The dashed line is the averaged SED and the dotted
line gives the observed 1$\sigma$ dispersion. Open circles are upper
limits. The average SED of spirals appears to be quite similar to that
of ellipticals and thus is not presented here.  The radio data do not
constrain the optical-MIR SED but are shown here for completeness. }
\end{figure}
\begin{figure}
\includegraphics[width=9cm,angle=0]{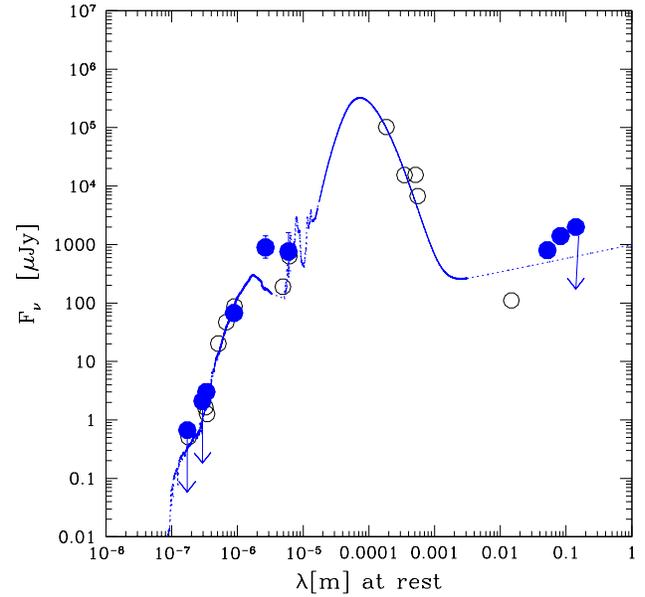}
\caption [] {Rest frame SED of \so\ (filled circles) compared with the
strong starburst HR10 (open circles), an ERO with a similar
redshift. The HR10 spectrum (Elbaz et al.\ 2001) has been normalised
by a factor 2.4 in order to match the observed $K$ magnitude of
\so. The line shows the best fit to the HR10 points assuming a pure
starburst model, constrained particularly in the MIR range by the
ISOCAM CVF dataset on local galaxies (Chanial 2001). No LW2 (5--8.5
$\mu$m) value is available for HR10; instead, LW10 (8--15 $\mu$m) was
obtained. Note that the well known weak 3.3 $\mu$m (rest frame) PAH
line marginally affects the LW2 band, whereas it is fully encompassed
by the LW10 band. }
\end{figure}
We now consider the $R$, $I$ \& $K$ photometry, which defines \so\ as
an ERO. With a $K$ magnitude of 17.5, \so\ is brighter than most EROs
but fainter than known quasars at a similar redshift (Wright et al.\
1983, Falomo et al.\ 2001).  Comparison with local templates (Fig.\ 3)
shows that, on the basis of these colours alone: (i) the pure
starburst hypothesis seems unlikely, and (ii) \so\ is compatible with
either an elliptical or a Seyfert 2 galaxy at a redshift of 1.5.
On the other hand, the ISOCAM data points clearly exclude a normal
elliptical galaxy. These two MIR values are likely to have different
origins: between 5--16 $\mu$m (rest frame), MIR emission in galaxies
is due mainly to black-body radiation from the photospheres of the
evolved stellar population and/or dust heated either by stars or an
active nucleus. The fact that in \so\ the 2.7 $\mu$m (rest frame) flux
is significantly higher than that of a standard Seyfert galaxy
suggests the presence of an additional hot component. Indeed, the
excess from 1--3 $\mu$m, can naturally be explained by very hot dust
emission in the circumnuclear region of an AGN (e.g. Kobayashi et al.\
1993). These grains can reach temperatures of a few thousand degrees
before they sublimate.
In addition, we have compared \so\ with another ERO, for which ISOCAM
photometry is also available (Fig. 4), namely HR10, ``a distant clone
of Arp 220" at $ z = 1.44$ (Elbaz et al.\ 2001).  In HR10 the bulk of
emission at 15 $\mu$m appears to be related to star formation rather
than to the presence of an active nucleus with a hot dust
component. Again, the two objects show very similar SEDs between the
observed $I$ and $K$ bands.  Furthermore, this comparison extends to
15 $\mu$m, but not around the 7.5 $\mu$m point which appears
significantly higher for \so.  This supports the previous evidence
that the emission process is much more energetic in \so, consistent
with the presence of a powerful central engine.
From the radio data, we find a total intrinsic power of
$P_{1.4\,{\rm GHz}}=1.9\times 10^{25}$ W~Hz$^{-1}$ (rest frame),
using the spectral index of $-$1 derived from the 20 and 13 cm
measurements. This places \so\ within the Fanaroff \& Riley (1974)
break range
$P_{1.4\,{\rm GHz}} \sim 10^{24}-10^{26.5}$ W~Hz$^{-1}$, shown to be a
strong function of the absolute isophotal magnitude of the galaxy by
Ledlow \& Owen (1996). Our current limit on the radio angular size of
\so\ is compatible with the source falling into the class of compact
steep-spectrum quasars.
The NIR spectrum of \so\ shows some obvious similarities to that of
the ultra-steep-spectrum red quasar WN J0717+4611 at $z = 1.462$ (de
Breuck et al.\ 1998), particularly in regard to the line ratios. The
latter object is polarized, and the authors argue that the origin of
the polarization is scattering by small dust grains or electrons.
Indeed, the presence of dust appears to be common in radio-loud
quasars, with the extinction being a function of the inferred viewing
angle to the radio axis (Baker \& Hunstead 1995).  \\
In conclusion, ISO's MIR view, combined with radio imaging and NIR
spectroscopy , has proven to be an efficient way of identifying a
dusty quasar at $z = 1.5$. Recent ISOPHOT results on PG quasars
have already revealed considerable dust emission between 25--200
$\mu$m (Haas et al.\ 2000). In \so\ ISOCAM provides evidence for
the presence of hot dust (at 2.7 $\mu$m, rest frame) heated by an
active nucleus. In this respect, \so\ appears to be a rare object.
With the present data it is not possible to gauge the magnitude of
a possible starburst contribution to the observed MIR
luminosity. Higher resolution optical, infrared and radio imaging may
enable a morphological study of \so, which may in turn shed light on
the origin of the active nucleus; so far, only 10\% of the ERO
population remain unresolved by HST (Stiavelli
2000). FIR/submillimeter observations would sharpen the SED picture,
revealing the presence of cooler dust --- responsible for the large
peak predicted by the model in Fig.\ 4 --- associated with star
formation activity.
\acknowledgements{We are grateful to F.\ Comeron and J.-G.\ Cuby for
help in preparing and performing the VLT/ISAAC observations presented
here.}
{}
\end{document}